# Patterns of cooperation during collective emergencies in the help-or-escape social dilemma

**Mehdi Moussaïd[1*] and Mareike Trauernicht[1]**

[1] Center for Adaptive Rationality, Max Planck Institute for Human Development, 14195, Berlin, Germany.
*Corresponding author: moussaid@mpib-berlin.mpg.de

# Abstract

Although cooperation is central to the organisation of many social systems, relatively little is known about cooperation in situations of collective emergency. When groups of people flee from a danger such as a burning building or a terrorist attack, the collective benefit of cooperation is important, but the cost of helping is high and the temptation to defect is strong. To explore the degree of cooperation in emergencies, we develop a new social game, the help-or-escape social dilemma. Under time and monetary pressure, players decide how much risk they are willing to take in order to help others.

Results indicated that players took as much risk to help others during emergencies as they did under normal conditions. In both conditions, most players applied an egalitarian heuristic and helped others until their chance of success equalled that of the group. This strategy is less efficient during emergencies, however, because the increased time pressure results in fewer people helped. Furthermore, emergencies tend to amplify participants' initial tendency to cooperate, with prosocials becoming even more cooperative and individualists becoming even more selfish. Our framework offers new opportunities to study human cooperation and could help authorities to better manage crowd behaviours during mass emergencies.

# Introduction

Mass evacuations in stressful emergency situations are extraordinary events during which people's behaviour is difficult to anticipate[1]. Recent examples include terrorist attacks in crowded urban areas (e.g. in Paris in 2015), crowd disasters during mass gatherings (e.g at the Mecca in 2015), and collective evacuations after natural disasters (e.g. the Ecuador earthquake in 2016). Understanding how people react when their lives are at stake is central to anticipating the collective behaviour of the crowd and helping authorities to manage such critical situations efficiently [2].

Numerous empirical analyses have been conducted after major incidents such as the 9/11 attacks[3], the Love Parade disaster[4], the crowd stampede at the Mecca pilgrimage[5,6], and several fire evacuations[7–9]. These studies have highlighted the role of various environmental, cognitive and social factors in facilitating or hindering mass evacuations, such as how the nature of the fire alarm affects people's reaction time, the preference for using familiar routes during egress, and the importance of social identification between individuals. Other experimental studies have demonstrated that people's risk perceptions are subject to social contagion, which can lead to the collective amplification or underestimation of the danger, potentially causing critical evacuation delays[8,10,11].

Cooperation in stressful emergency situations is known to have important collective benefits: First, cooperation reduces the frequency of pushing and competitive behaviours[12], typically associated with flow disturbances around exit doors[13,14]. Second, cooperative individuals can help people who have fallen in a rushing crowd back to their feet[12], thus reducing the risk of stampedes and avoiding the obstructions caused by the bodies on the ground[4,13]. Third, collective solidarity can mitigate fear and negative emotions, thus reducing the risk of panic. Fourth, individuals sharing their private routing information can help others find their way out and avoid detrimental herding patterns towards dead ends or unusable doors[15].

However, the extent to which people behave altruistically in emergency situations remains unclear. Early research indicates that people's behaviour under emergency conditions is rather egoistic. For example, the mass panic approach describes individuals as acting in a purely selfish manner[16,17], and Mintz's seminal experiment suggests that individual's cost-benefits calculations lead to jams at bottlenecks and inefficient evacuations[18]. In contrast, surveys conducted among emergency survivors point to a high level of cooperation within groups of individuals facing a deadly threat[19]. Furthermore, laboratory experiments and empirical analyses of real emergencies have shown that solidarity is increased when there is a strong feeling of social identification between individuals[5,12,20]. A recent virtual evacuation experiment showed that a considerable proportion of participants tend to behave altruistically during emergencies, but that proportion diminishes as the cost of helping increases[21].

Thus, the literature on mass evacuations does not paint a clear picture of how much cooperation can be expected during emergencies and which social and environmental factors promote solidarity within the group. However, collective emergencies share characteristics with social dilemmas, on which there is an extensive literature. Social dilemmas are defined as situations where individual and collective interests yield opposed behaviours: the whole group is better off if each individual cooperates, but each individual has more gain from defecting[22,23]. Research on social dilemmas and human cooperation more generally suggests opposing hypotheses with regard to collective emergencies. On the one hand, the social heuristic hypothesis suggests that fast and intuitive decisions made under time pressure promote cooperation, in contrast to calculated behaviours deriving from an internal deliberation process[22,24]. The severe time pressure typical of emergency situations should thus boost solidarity and prosocial behaviours within a crowd. On the other hand, emergency situations are characterized by the nature of the underlying reward structure, which is, by definition, in the loss domain[25]: In situations of life and death, people compete to avoid an important loss – that of their life, health, or valuable possessions. Escaping safely does not provide any extra bonus. According to the prospect theory, people are generally loss averse[26] and should have a stronger urge to escape, and therefore a smaller tendency to help others when their lives are at stake. In line with this assumption, Mather and Lighthall have demonstrated that stress leads to a reduction in the choice of negative outcomes[27]. But prospect theory does not take the influence of time pressure and the involvement of other people into account, and the social heuristic hypothesis makes no exact claim about the influence of losses or gains.

The challenge of understanding cooperation under emergencies is that multiple, co-occurring factors could have opposing effects. A simple experimental and conceptual framework is therefore needed to explore cooperation in emergencies in detail, in the same way as economic games have helped researchers to explore other social dilemmas[28].

For ethical reasons, it is clearly not possible to expose subjects to actual emergency situations in the laboratory. However creating conditions that closely mimic important aspects of emergency situations is conceivable. To this end, we have designed a new social game called the help-or-escape social dilemma, which reproduces the core features of collective emergencies. In the next sections, we formally describe this framework and present first experimental results describing how experimental subjects behave in it. We show that individual levels of cooperation in the game are in line with other standardized measures of altruism. Our findings indicate that participants react differently to increased pressure, with prosocial players tending to show increased cooperation during emergencies and individualists tending to become even more selfish.

## The "Help-or-Escape" Game

The help-or-escape (HoE) game is a conceptual and experimental framework designed to mimic the social dilemmas that typically occur during collective emergency situations. The game puts subjects in a situation where they have to decide how much risk they are willing to take in order to help other participants.

The HoE game operates under time pressure. That is, the outcome of the game depends not only on the subject's decisions, but also on the timing of those decisions. In each instance of the game, the subject has a probability $p(t)$ of receiving a bonus $\alpha \geq 0$ and the complementary probability $(1 - p(t))$ of receiving a penalty $\beta \leq 0$. The bonus $\alpha$ represents the gain associated with success (i.e., managing to escape), whereas the penalty $\beta$ represents the cost associated with failure (i.e., not managing to escape). The probability $p(t)$ is time dependent with $p(0) = 1$ and $p(\tau) = 0$, where $\tau$ is the duration of the game. Assuming that $p(t)$ decreases linearly with time, we have the probability decay $\gamma = 1/\tau$ per unit of time, and thus $p(t) = 1 - \gamma t$, for $t$ ranging from 0 to $\tau$.

At any moment, the subject can decide to end the game and leave the setting (i.e., to escape). From the time $t$ when that decision was made, we derive the probability $p(t)$, which is then used to determine whether the subject receives the bonus $\alpha$ (i.e., success) or the penalty $\beta$ (i.e., failure).

The above design describes a simple environment with time pressure: An immediate decision to leave the setting almost guarantees success, whereas a late decision has a higher chance of failure. Time pressure can be easily manipulated by varying the total time $\tau$; whereas the values of $\alpha$ and $\beta$ enable a control of the risk magnitude. For example, a large value of $\tau$, $\alpha > 0$ and $\beta = 0$ describes an environment with little time pressure, where the individual is rewarded for success and does not risk any penalty in case of failure. An emergency situation, in contrast, would be best described by a small value of $\tau$, $\alpha = 0$ and $\beta < 0$, where time pressure is strong and the subject seeks only to avoid a loss.

In the absence of any other constraints, subjects would immediately leave the setting at $t = 0$, maximizing their chance of success. The social dilemma arises from the presence of other individuals whom the focal subject can decide to help before leaving the setting. The social environment is composed of N other players who have a fixed low probability $p_h$ of succeeding and receiving the bonus $\alpha$, and a fixed high probability $(1 - p_h)$ of failing and receiving the penalty $\beta$. These participants are passive players: they make no decisions and can only wait and hope to be helped by the focal subject. They represent individuals in need whom the focal subject meets during the escape, such as injured people asking for help during a fire evacuation. At any moment before leaving the setting, the focal subject can decide to help one of these participants. The success probability $p_h$ of that person being helped is then increased to $p'_h$. The probabilities $p_h$ and $p'_h$ are game parameters, with $p'_h > p_h$. The act of helping costs time, however, and therefore reduces the helper's chances of success. In our game, the decision to help another person disables all other possible actions during the helping time (i.e. the subject cannot leave the setting or help another person), which corresponds to the loss of $p_c$ chances to escape successfully. Therefore, the focal subject can offer another person a higher success chance of success *at the expense* of his or her own chance to escape. Once the helping time is over, the subject is free to leave the setting or to help another person. Because the available time decreases continuously, hesitations and thinking delays also reduce the subject's chances of success. The game ends once the subject decides to leave the setting or when the total time $\tau$ is over. At the end of the game, the focal subject and the *N* other players have success probabilities that depend on the sequence of decisions made by the focal subject, from which every participant's gain or penalty can be determined.

The game enables various factors possibly influencing cooperation to be evaluated, such as time pressure, risk magnitude, the number of people in need, the utility of the help provided, and the cost of helping. **Table 1** summarises the parameters of the game. Note that helping others is not rewarded in the game: the decision to help can thus arise only from altruistic motivations. This framework describes the structure of a variety of situations with different levels of emergency. In the next section, we present two situations with distinct levels of emergency, and experimentally test subjects in them.

## Experimental procedure

In order to feed our conceptual framework with empirical data, we implemented two distinct environments and experimentally tested a total of **104** participants in both environments. In the *baseline* condition, we tested how much help was given by the focal subject in a situation of low emergency; in the *emergency* condition, we tested the same subjects in a situation of high emergency. A third, intermediate condition was also implemented but the results were identical to those of the baseline condition and are thus not reported in the present article (see the Supplementary Information for the results of this intermediate condition).

In each condition, we exposed subjects to an introductory framing story to establish a clear reference to reality. The baseline condition described an everyday life situation with little time pressure and only minor consequences in case of success or failure. Specifically, we asked subjects to imagine the following scenario: They are at a train station and want to catch a certain train. On their way to the platform they meet other individuals who need their help finding their own train. If the subjects decide to help one or more of the individuals in need, their own chances of catching their train decrease, whereas the chances of individuals they have helped increase. To emphasize that it was a low-stress situation, we told subjects that they had arrived early at the train station and informed about the route to their platform. There was therefore no major emergency involved. We set $\tau = 60$ seconds, $\alpha = 1$ euro, and $\beta = 0$. Therefore, the risk involved is defined in the *gain* domain, that is, subjects are trying to receive a bonus, with no penalty in case of failure[25].

The emergency condition, in contrast, involved severe time pressure and major consequences in case of failure. Specifically, we asked subjects to imagine the following: They are at a train station and an explosion occurs, causing parts of the building to come crashing down. They need to leave the building as soon as possible. As in the baseline condition, they meet other individuals in need on their way out. Helping decreases their own chances of making it out of the building on time, but increases the chances of the people they helped to avoid an important loss. To emphasize that it was a high-stress situation, we told subjects that they did not know which routes were still passable. The major emergency was represented by the parameters $\tau = 15$ seconds, $\alpha = 0$, and $\beta = -4$ euros. Unlike the baseline condition, the risk here is defined in the *loss* domain, that is, subjects are trying avoid an important loss, with no monetary bonus in case of success [25].

For both conditions, we set the other parameters to $N = 8$, $p_h = 0.1$, $p'_h = 0.6$, and $p_c = 0.1$. We choose $N = 8$ to give the subject a chance to help all people in need, which would cost 8x$p_c$, that is, 80% of their chance of success plus some response time. The values of $p_h$ and $p'_h$ indicate that the chance of success of people who receive help increases from 10% to 60%. We chose 10% to clarify that there was a chance that the person would be fine without help, but that it was very low. We chose 60% to indicate that, with help, a positive outcome for the helped person was more likely than by chance, and that helping therefore had a considerable effect.

We ran a series of five rounds of the baseline condition with each subject, followed by a series of five rounds of the emergency condition (see **Materials and Methods**). The two conditions were presented to subjects in the same order. The purpose of this fixed-order design is to control for order effects: People facing emergency situations after a series of baseline situations might react differently than when the order is reversed. In order to stay as close as possible to the naturally occurring sequence of cooperation in daily-life situations, we chose to study situations where the emergency condition follows a series of baseline situations, rather than the opposite. As a side effect, our results are only valid in situations where this order of events apply, or under the assumption that the order effect is negligible. In addition, we measured the subjects' social value orientation (SVO) — a standardized measure of the magnitude of the concern people have for others. We used the newly developed (SVO) Slider Measure described by Murphy et al.[29], a six-items resource allocation task distinguishing altruistic, prosocial, individualistic, and competitive tendencies. The focal subject's decisions in the game and the results of the SVO test affected *real* other participants who actually received a monetary bonus or penalty depending on the focal subject's decisions. Subjects were informed of this payment scheme in advance; the behaviour we observed therefore had real and not just hypothetical consequences.

# Results

We first evaluated variability of behaviours across the five rounds of each condition. A one-way ANOVA indicated no significant differences across the rounds with regard to the number of people helped in the baseline condition, $F(4, 515) = 0.04$, $p = .99$, or the emergency condition, $F(4, 515) = 0.17$, $p = .95$. We therefore used the mean value of help provided over the five rounds for each subject and condition in all subsequent analyses.

**Aggregate level.** The degree of cooperation was evaluated by means of two different measures, which we call the outcome level and the process level. At the outcome level, cooperation $C_n$ is measured by looking at the *number* of participants in need who were helped by the focal subject. At the process level, cooperation $C_p$ is evaluated by looking at *how much risk* the subjects took to help others, that is, $C_p = 1 - p(t_x)$, where $t_x$ is the time at which the focal subject left the setting. The outcome and process levels are not exactly equivalent: For example, a subject may want to help five individuals no matter how much risk is involved, or to exit the setting when their chance of success equals 50%, no matter how many people have been helped. The number of people helped (outcome) does not always match the risk taken (process), especially if too much time has been spent on motor actions or lost to thinking delays and hesitation.

**Figure 1** shows the cumulative probability distributions for all experimental subjects at the outcome and process levels, and for both conditions. Subjects helped significantly fewer people in the emergency condition than in the baseline condition (a two-sample *t*-test revealed a statistically significant difference with p = 0.034). In the baseline condition, subjects helped on average 3.71 (SD = 2.11) people, and in the emergency condition, 3.11 (SD = 2.01) people. At the process level, however, no difference was detected (p = 0.93; two-sample *t*-test). On average subjects in the baseline condition gave 42.8% (*SD*=22.56) of their chance of success to help others; subjects in the emergency condition gave 43.1% (*SD*=25.35).

What is the origin of this discrepancy between the outcome level (subjects help fewer people) and the process level (subjects take the same risk to help others)? As we hypothesized earlier, this difference can be explained by processing time. **Figure 2** shows the absolute and effective processing time under both conditions. The processing time is defined as the total time spent in the setting during which the subject was *not* helping another person. It covers delays caused by the motor actions of moving the mouse and clicking on buttons, as well as thinking and hesitation time. The *absolute* time is measured in seconds, whereas the *effective* time is measured in terms of lost chances to succeed and corresponds to the absolute time multiplied the probability decay $\gamma$. As shown in **Figure 2**, subjects actually reacted faster and/or spent less time thinking about their next action in the emergency condition (black bars). However, this increased processing speed was not sufficient to compensate the increased time pressure of the environment. In fact, the effective time has almost doubled in the emergency condition (**Figure 2**).

From this first series of results, we can picture the mechanisms operating at the aggregate level: On average, subjects gave approximately 40% of their chance of success to help other participants and tried to apply this strategy in *both* environments. In the baseline condition, participants needed on average 3.5 seconds for the processing time corresponding to 6% of their chance (because the probability decay is 1.5% per second in this condition). Therefore, among the 42.8% chance that participants gave for cooperation, only 36.8% were *effectively* used for helping. Consistently, participants managed to help on average 3.71 persons in this condition. In the emergency condition, participants maintained the same intention to cooperate: they gave 43.1% of their chance to help others and used on average 1.8 seconds for the processing time. However, the probability decay is faster in the emergency condition (i.e., 6.5% per second). Consequently, their processing time corresponded to 12% of their chance. Therefore, among the 43.1% chance that participants gave for cooperation, only 31.1% were effectively used for helping. Consistently, participants only managed to help 3.11 persons in emergency. The results of the intermediate conditions shown in the supplementary information also exhibit the same pattern. Therefore, it *appears* that subjects helped less in the emergency condition, but this difference was mostly due to changes in the environmental conditions rather than to an actual decay of the helping intentions.

The fact that subjects left the setting with a 60% chance of success (i.e., gave 40% of their chance of success to help others) coincides with the game parameter $p_h = 0.6$, corresponding to the success probability of helped individuals. The most likely interpretation of this similarity is that subjects tended to apply an *egalitarian* strategy, whereby they were willing to cede part of their initial chance of success to align it with the group's probability of success. In other words, subjects helped as many others as they could until their own chance of success was equal to that of the people they helped.

**Individual level.** In addition to aggregate trends, we examined behavioural changes at the individual level. Personal SVO scores were calculated according to the procedure described in [29]. On average, subjects scored 30.37° (*SD* = 12.16). Based on their scores, subjects were categorized into one of four groups defined in Murphy et al.[29]: 1 subject was classified as having an altruist profile (defined as SVO° > 57.15); 77 subjects, a prosocial profile (defined as 22.45 < SVO° < 57.15); 26 subjects, an individualistic profile (defined as -12.04 < SVO° < 22.45), and none a competitive profile (defined as SVO° < -12.04). As only one subject was categorized as an altruist and none as a competitor, we analysed group differences in cooperation between subjects with a prosocial and an individualistic profile (independent-samples *t*-test). Cooperation in the low-stress conditions was significantly higher for prosocial individuals (*M* = 4.31, *SD* = 1.91) than for individualists (*M* = 1.99, *SD* = 1.71), *p* < .001. The same applied to the emergency condition (prosocials: *M* = 3.72, *SD* = 1.87; individualists *M* = 1.35, *SD* = 1.24; *p* < .001). The results of our experiment are therefore consistent with the standardized measure of SVO, indicating that our setup captures some basic components of human cooperative behaviour.

We then examined change in individual subjects' cooperation between the baseline and emergency conditions. For this, we focus on process level cooperation, that is, how much of their success chance they gave for helping others. In such as way, we highlight changes in the participants' helping intentions, controlling for the side effects of the increased time pressure. To this end, we measured the normalized change $\delta$ between cooperation at the process level $C_p^0$ in the baseline condition and cooperation at the process level $C'_p$ in the emergency condition, for each individual:

$$\delta = (C'_p - C_p^0) / d$$

Here, $d$ is a normalization factor that is set to $d = C_p^0$ if $C'_p < C_p^0$, and $d = (1 - C_p^0)$ if $C'_p > C_p^0$. The normalization reflects the fact that the change $\delta$ is measured relatively to how much change is possible given the baseline level of cooperation $C_p^0$. **Figure 3** shows the observed values of $\delta$ for all subjects as a function of their baseline cooperation $C_p^0$. As expected, the split between the individualists (in red) and the prosocials (in blue) is clearly visible along the x-axis: the majority of individualists exhibited a lower cooperation level than the prosocials. The figure also reveals an important variability in how each individual reacted in the emergency condition. Among the subjects who provided little help in the baseline condition, some provided more help in the emergency conditions (quadrant A), while others provided even less (quadrant C). Likewise, among the subjects who provided a lot of help in the baseline condition, some provided less in emergency (quadrant D), while others provided even more (quadrant B). Overall, however, the baseline cooperation level $C_p^0$ significantly correlates with the magnitude of change $\delta$ (R = 0.24, p = 0.01, irrespective of the SVO category). That is, subjects who provided more help in the baseline condition tended to increase their help under emergency, while those who provided little help in the baseline condition tended to decrease their help under emergency. Hence, the emergency condition tended to *amplify* the subjects' initial cooperative tendencies. As **Figure 4** shows, this trend is also visible when subjects were grouped according to their SVO profiles. In emergencies, the majority of individualists (52%) reduced their cooperation level, while 44% of the prosocials increased theirs (**Figure 4A**). Furthermore, a Kruskal-Wallis test comparing the full distributions of change $\delta$ between prosocials and individualists yields a p-value of .0046 confirming the above tendency (**Figure 4B**). This emergency-induced amplification of the cooperation profiles was not detectable at the aggregate level due to the moderate intensity of this tendency combined with the fact that the changes of prosocials and individualists tended to cancel each other out at the global level.

## Discussion

The help-or-escape game is a multi-player framework in which subjects decide how much risk they are willing to take in order to help other participants in anonymous, unidirectional, one-shot interactions. The game mimics social dilemmas that typically occur in collective emergency situations, using probability rates instead of fixed numbers to indicate expected outcomes. The game parameters allow a variety of situations to be implemented and various factors potentially influencing cooperation to be evaluated, such as time pressure, risk magnitude, group size, utility of the help provided, and cost of helping. In the present definition of the framework, we chose to keep the structure of the game as simple as possible. Nevertheless, numerous extensions can be considered in future improvements. For example, competitive behaviours could be implemented, allowing subjects to harm other participants to increase their own chance of success; alternatively, several players could be active at the same time, opening up a range of issues related to bystander effects, social influence, and diffusion of responsibility with regard to helping people in need[30,31].

As a starting point, we examined behavioural differences between a baseline condition imitating situations often encountered in daily life, and a high-risk situation imitating exceptional emergency events. The main differences between the two conditions were (1) the increased time pressure, and (2) the nature of the risk, which changed from the gain to the loss domain. We purposely exposed our participants to the baseline condition first followed by the emergency condition after. The order in which the conditions were presented was identical for all the participants. The purpose of this fixed-order design was to control for possible order effects: People exposed to an increasing level of emergency might react differently than those exposed to a decreasing level of emergency. For example, people who experience high-stress situations in their daily life, such as populations living in war zones, might develop cooperation strategies that are adapted to the situations they encounter most frequently, and thus react differently to environmental changes. To control for this hypothetical order effect, we chose to study only the specific situations where emergency situations follow a series of non-emergency situations, because this order is arguably closer (although not identical) to daily-life situations than the opposite order. Our results and conclusions, therefore, are only valid given this specific order of events, or assuming a negligible order effect. Beside, we provide additional results in the supplementary information showing that habituation effects did not affect our experimental results (**Figure S2**).

Our results show that subjects tended to help less in the emergency condition than in the baseline condition. However, this decrease was not due a reduced intention to cooperate. In fact, subjects took as much risk to help others in both conditions, but the increased time pressure in the emergency condition did not allow them to help as many others as in the baseline condition. The amount of risk that subjects took in each condition (approximately 40%) suggests that they were relying on an egalitarian heuristic[33]. That is, they helped others until their own chances of success were equal to those of the people they were helping.

This mechanism can produce counterintuitive results. An individual may maintain the same helping intention in the emergency condition or even increase it, but end up helping fewer people. For instance, suppose that an individual is willing to give 40% of his or her success chance for helping others, and needs 2 seconds of processing time. In the baseline condition, the time pressure is low and the probability decay is approximately 1.5% chance per second. Therefore, 2 seconds of processing time correspond to a loss of 3% success chance. Among the 40% chance that the individual is willing to give, 3% are lost and 37% will be effectively dedicated to help, which correspond to an average of 3.7 persons helped. In the emergency condition, however, the 2 seconds of processing time correspond to a loss of 12% success chance because the probability decay rate is approximately 6.5% chance per second in this condition. Thus, among the fixed 40% chance that the individual is willing to give, only 28% will be effectively dedicated to the help, which correspond to an average of 2.8 persons helped. Therefore, the same intention to help in both conditions results in fewer people being helped in the emergency condition. In fact, the individual would need to increase his or her helping intention up to 52% only to counteract the effect of the increased time pressure. In our experiment, participants tended to maintain the same intention to help across conditions at the aggregate level, which resulted in fewer people being helped. The observed decay of cooperation thus results from a mechanical side effect of the increased time pressure, rather than behavioural change[32].

In the literature on human cooperation, the social heuristic hypothesis suggests that intuitive and fast decisions promote cooperation [24]. People are therefore expected to cooperate more when time pressure increases, which is not in line with our observation in the HoE game at the aggregate level. However, time pressure is not the only factor that changes in emergencies. Numerous other features of the environmental structure are also affected, ranging from the global context (represented by the framing story we told our subjects), to the nature of the risk, which switches from the gain to the loss domain. In this respect, prospect theory states that people generally tend to avoid high losses[26], which would instead lead to a decrease in cooperation in our setup. It is therefore difficult to map theoretical predictions onto concrete emergency situations due to the large number of co-occurring environmental factors. The HoE game makes it possible to decompose these multiple influencing elements, and thus to understand the complex mechanisms operating in emergency situations.

Although the helping intention remains unchanged at the aggregate level, our results show that individuals responded to emergencies in different ways offset each other at the aggregate level. In particular, we found that individual cooperation profiles tended to become more extreme under emergency. This observation is partly in line with a recent in-depth study of spontaneous cooperation[34], which demonstrated that prosocial individuals increased their level of cooperation under time pressure, but individualists did not. Our findings confirm these results and also show the symmetric effect, namely, that individualists reduce their level of cooperation in emergencies. Emergencies therefore seem to amplify people's natural cooperation tendency. This interpretation of our results is generally consistent with the picture emerging from recent research on human cooperation[22,23].

In conclusion, the help-or-escape framework offers new research opportunities and can be used to experimentally address numerous issues related to cooperation in collective emergencies. The HoE game can therefore help to advance research on crowd management and issues related to human cooperation in general.

## Materials and Methods

**Experimental procedure**. Data were collected in Berlin, Germany, between July 1$^{st}$ and July 31$^{st}$, 2015. A total of **104** participants were recruited for the study (47% male, $M_{age}$ = 24.9 years). The study was approved by the Ethics Committee of the Max Planck Institute for Human Development. All methods were performed in accordance with the approved guidelines. All participants gave informed consent to the experimental procedure. The experiment was computer-based; participants were seated in front of a computer screen for the duration of the experiment. Each participant played the role of the focal subject for 5 repetitions of each of the three following conditions: the baseline condition ($\alpha = 1; \beta = 0; \tau = 60s$), an intermediate condition ($\alpha = 0; \beta = -1; \tau = 30s$), and the emergency condition ($\alpha = 0; \beta = -4; \tau = 15s$). The three conditions were presented in this order. The results from the intermediate condition did not differ from those of the baseline condition and are not reported in this article. The other parameters of the game remained unchanged across conditions, as shown in **Table 1**. For each condition, we first exposed the subject to a framing story and explained the rules of the game. After correctly answering three comprehension questions, subjects did a practice trial and finally played 5 consecutive rounds with the same game parameters. This process was repeated for all three conditions. After the third condition, subjects entered some personal demographic information and completed a Social Value Orientation (SVO) measurement using the SVO Slider test described in Murphy et al.[29]. In that test, participants made six unilateral decisions about the allocation of resources between themselves and another person. In each of the six items, participants were confronted with a continuum of payoffs for themselves ranging from 50 to 100 cents. Each of these payoffs was associated to a payoff for another person ranging from 15 to 100 cents. Participants had to choose which pair of payoffs they prefer. The exact payoff values for each item were determined and calibrated in Murphy et al.[29]. Finally, subjects were paid and dismissed.

**Payment**. Participants received a starting capital of 14 euros at the beginning of the experiment, and were informed that this amount would increase or decrease depending on the outcome of the experiment. Participants were explicitly informed that their chance to receive the bonus $\alpha$ and avoid the penalty $\beta$ in each round decreased continuously as long as they did not decide to escape by pressing a button. At the end of the experiment, we randomly selected one round from each condition, and one decision in the SVO test. The total monetary payoff for those selected items was then determined and added to the subject's starting capital. This could result in an increase or a decrease of the starting capital, depending on the decisions made by the subject. The subject also passively played the role of one of the $N$ individuals in need of the previous experimental session. That is, we randomly selected one of the $N$ individuals in need for each condition of subject $i - 1$, and transferred the corresponding payoff to the capital of subject $i$. Thus, the decision made by each participant affected the outcome of the next one. Participants were informed about this procedure. The payoff inherited from the previous participant was only made known to the subject at the very end of the experiment to ensure that the subject's behaviour was not influenced by direct or indirect reciprocity.

**Experimental software**. In each round, time was symbolized on the screen by an arrow that continuously moved down a vertical scale ranging from 100% (top) to 0% (bottom). The speed of the arrow was calibrated on the value of $\tau$ in each condition such that participants had a visual representation of the time pressure. In addition, in the emergency condition, a red blinking frame was added to the screen to emphasize the alarming atmosphere. The number of participants in need $N$ was represented by a set of $N$ stylized illustrations of people. Every time the subject decided to help someone, one of these illustrations was automatically removed from the set to show how many more people could be helped. Subjects indicated their decision to help or to leave the setting by clicking on a respectively labelled button. When subjects decided to help, both buttons were disabled during the helping time, corresponding to a loss of $p_c$ chances to succeed.

# Acknowledgements

We are grateful to Susannah Goss and Dania Esch for editing the manuscript. We thank Monika Keller, Thorsten Pachur, Tyler Thrash, Mubbasir Kapadia, and Ryan Murphy for fruitful discussions. This research was supported by a grant from the German Research Foundation (DFG) as part of the priority program on *New Frameworks of Rationality* (SPP 1516) awarded to Ralph Hertwig and Thorsten Pachur (HE 2768/7-2). The funders had no role in study design, data collection and analysis, decision to publish, or preparation of the manuscript.

# Contributions

M.M. and M.T. designed the research, conducted the experiment, analysed the data, and wrote the paper.

# Competing interests

The authors declare no competing financial interests.

# Figure legends and tables

| Parameter | Description | Baseline condition | Emergency condition |
|---|---|---|---|
| $\alpha$ | Success bonus | +1 | 0 |
| $\beta$ | Failure penalty | 0 | - 4 |
| $\tau$ | Duration of the game | 60 seconds | 15 seconds |
| $N$ | Number of participants in need | 8 | 8 |
| $p_h$ | Success probability for non-helped participants in need | 0.1 | 0.1 |
| $p'_h$ | Success probability for helped participants in need | 0.6 | 0.6 |
| $p_c$ | Cost of helping for the focal subject | 0.1 | 0.1 |

**Table 1: Description of the game parameters** and values implemented in the baseline and the emergency conditions.

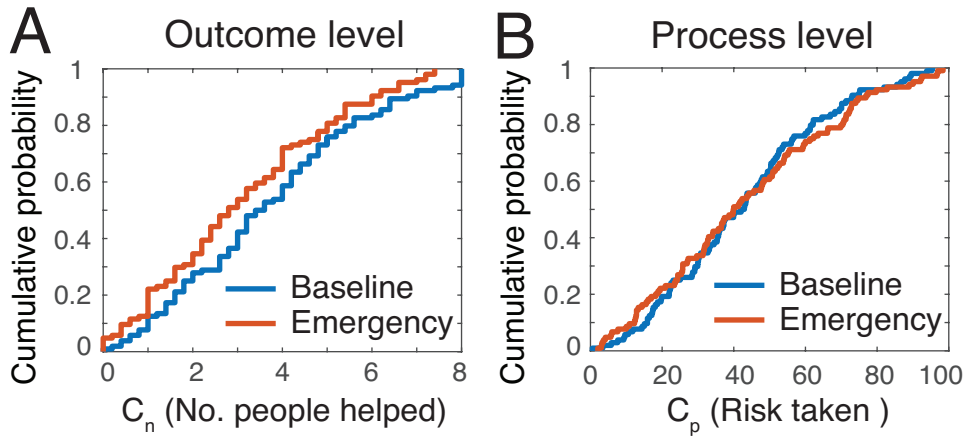

**Figure 1**: **Experimental results**. Cumulative probability distributions of the degree of cooperation observed in the baseline condition (in blue) and the emergency condition (in red), for all 104 participants. In (A), cooperation $C_n$ is measured at the outcome level, that is, in terms of how many participants the subject helped before leaving the setting. In (B), cooperation $C_p$ is measured at the process level, that is, in terms of how much risk was taken to help others (irrespective of how many people were helped). While no significant difference was observed at the process level, the level of cooperation was significantly lower under emergency conditions at the outcome level.

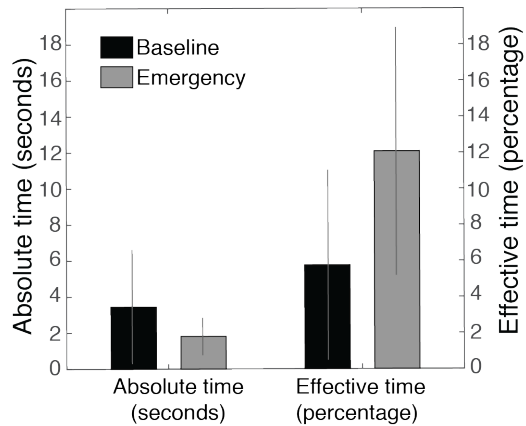

**Figure 2**: **Processing time**. Average time spent in the game setting without helping anybody under baseline (in black) and emergency (in grey) conditions, for all 104 participants. The absolute time is measured in seconds, whereas the effective time is measured in terms of lost chances to escape. Errors bars indicate the standard deviation of the mean. While subjects tended to react faster in the emergency condition, the effective processing time doubled. Chances decreased at a rate of approximately 1.5% per second under baseline conditions, and 6.5% per second under emergency conditions.

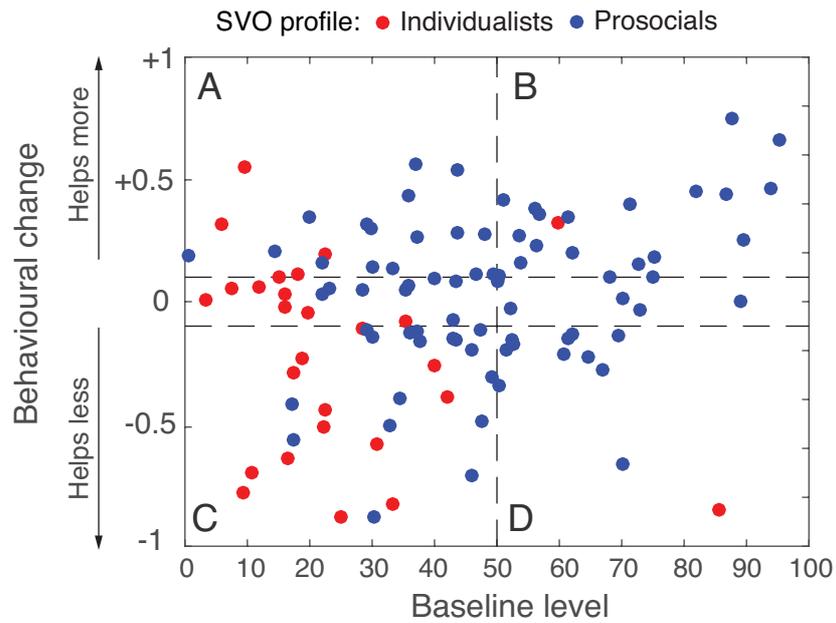

**Figure 3**: **Emergency-induced behavioural changes**. Individual changes in cooperation between the normal and the emergency conditions for all 104 participants. The baseline level represents how much help was given under normal conditions; the behavioural change indicates whether that subject gave more or less help in the emergency condition (normalized by the amount of possible increase or decrease). Red and blue dots indicate subjects classified as individualists and prosocials, respectively, as measured by their SVO. The quadrants (A-D) indicate whether subjects initially helped little (A,C) or a lot (B,D), and whether subjects increased (A,B) or decreased (C,D) their help under emergency conditions.

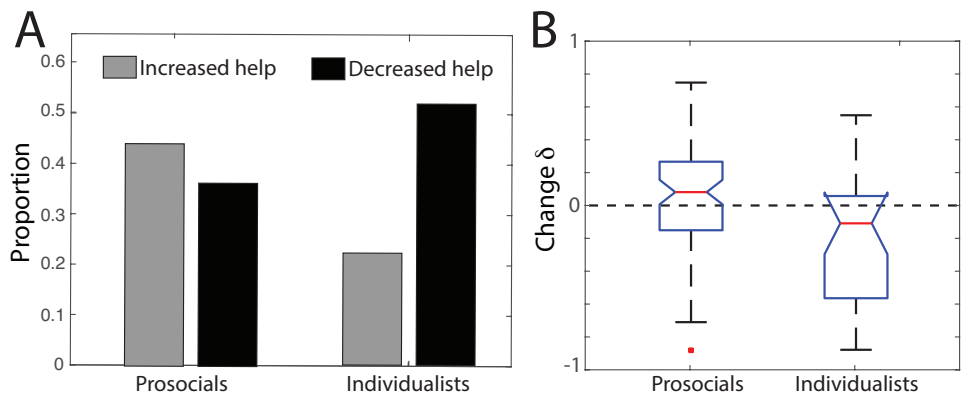

**Figure 4**: Behavioural changes induced by the emergency condition among the participants classified as prosocials (n=77) and individualists (n=26). (A) The grey bars indicate the proportion of individuals in each category who increased their help in the emergency condition ($\delta > 0.1$), and the black bars indicate the proportion of those who decreased their help ($\delta < -0.1$). The proportion of participants who did not change their help ($-0.1 \leq \delta \leq 0.1$) is not shown. The majority of prosocials (44%) increased their help during emergency whereas the majority of individualists (52%) decreased their help. (B) Boxplots comparing the full distributions of change $\delta$ for the prosocials and individualists. A Kruskal-Wallis test indicates a significant difference between prosocials and individualists (p = .0046).